\documentstyle{article}
\title{Quantization of the scalar field
in a static quantum metric}
\author{ Z. Haba\\Institute of Theoretical Physics, University of Wroclaw,
\\50-204 Wroclaw, Plac Maxa Borna 9, Poland\\e-mail:zhab@ift.uni.wroc.pl}
\date{}
\begin{document}
\maketitle
\begin{abstract}
We investigate Hamiltonian formulation of quantum scalar fields in
a static quantum metric. We derive a functional integral formula
for the propagator. We show that the quantum metric substantially
changes the behaviour of the scalar propagator and the effective
Yukawa potential.

\end{abstract}
\section{ Introduction}
There is an old idea coming from Landau, Pauli and others ( see
the review of the problem in 1957 by Deser \cite{deser}) that
quantization of gravity could cure the ultraviolet divergencies in
quantum field theory of matter fields interacting with gravity.
Since then this inspiring idea has been followed by other authors
(let us mention \cite{salam}-\cite{ford}). However, in general we
find their approximate methods unreliable. In our earlier papers
\cite{hab1}-\cite{habajmp} we have investigated an interaction of
a scalar field with a quantum or random metric in a  functional
integral formulation using the proper time representation for the
propagator. The interaction of gravity with the scalar field has
been treated non-perturbatively. We explicitly calculated an
average over the metric field. For the reason of a mathematical
consistency the functional measure of the metric field did not
satisfy conditions required for the quantum field. In such a case
it remained unclear whether the results could have been derived in
a unitary framework of Hamiltonian dynamics. In this paper we
study a similar model in the Hamiltonian formulation and relate
the results to the ones of the functional integral form. We make
an assumption that the metric does not depend on some coordinates
 in order to be able to perform explicit calculations. In particular,
 this assumption can be realized by a choice of a static metric. Then, we
discuss in detail the model of a scalar field interacting with a
static quantized metric. We are interested in a model where the
quantum metric comes from a time-zero quantum field. In general,
such a field evolves with time. In order to simplify the model we
neglect this time evolution. A perturbation of the free
Hamiltonian by a quantized metric is a singular perturbation
problem. So, we first start with a regularized metric which is an
analytic function. We discuss the operator Hamiltonian formalism
in the Fock space. We express the expectation values  of the
time-ordered products of the scalar field in the Fock space at
some special complex points by means of a rigorous version of the
Feynman integral.
 Then, the average over the metric field is calculated.
Finally, the regularization of the metric field can be removed. We
discuss in some detail an example of the metric being a
threedimensional free quantum field. We show that an interaction
with such a quantum metric substantially changes the short
distance behaviour of the scalar propagator. The Hamiltonian
approach allows to calculate the S-matrix and to calculate the
effective two-body potential. We discuss these problems briefly.
\section{The Hamiltonian approach}
We consider the usual Lagrangian for the scalar field
\begin{equation}
L=g^{AB}\partial_{A}\phi\partial_{B}\phi
\end{equation}
We assume  that the metric $g$  in $D$ dimensions  (with the
decomposition $D=d +(D-d)$ of the coordinates $x=({\bf
x},X)$)depends only on $d$ spatial coordinates ${\bf x}$ and is of
the block diagonal form $(g^{AB})=(g^{jk}({\bf x}),g^{\mu\nu}({\bf
x}))$ where $g^{AB}=\eta^{AB}+h^{AB}$ and $\eta^{DD}=-1$ with the
time $t=X_{D}$ (we shall concentrate here on the case $D=4$ and
$d=3$). Let $f_{\alpha}\exp(-i\omega_{\alpha}t)$ be the set of
positive frequency solutions (it is clear what it means because
the metric is time-independent) of the wave equation
\begin{equation}
-2{\cal A}\phi\equiv\partial_{A}(g^{AB}\sqrt{g}\partial_{B})\phi=0
\end{equation}
where $g=-\det(g_{AB})$.

We choose the normalization \begin{displaymath} \int d{\bf x}
\overline{f}_{\alpha}f_{\beta}=\frac{1}{\omega_{\alpha}}\delta_{\alpha\beta}
\end{displaymath}
We expand the solution $\phi$ into a complete set of positive
$f_{\alpha}\exp(-i\omega_{\alpha}t)$ and negative energy solutions
$\overline{f}_{\alpha}\exp(i\omega_{\alpha}t)$
\begin{equation}
\phi(t,{\bf x} )=\sum_{\alpha}f_{\alpha}({\bf
x})a_{\alpha}\exp(-i\omega_{\alpha}t) +\overline{f}_{\alpha}({\bf
x})a_{\alpha}^{+}\exp(i\omega_{\alpha}t)
\end{equation}
where $(a_{\alpha},a_{\alpha}^{+})$ are the usual annihilation and
creation operators in a Fock space ${\cal F}_{sc}$
\begin{equation}
[a_{\alpha},a_{\beta}^{+}]=\delta_{\alpha\beta}
\end{equation}
The Hamiltonian describing the time evolution (3) reads
\begin{displaymath}
H=\sum_{\alpha}\omega_{\alpha}a_{\alpha}^{+}a_{\alpha}
\end{displaymath}
 Let $\vert 0\rangle_{s}$ be the Fock vacuum for
the scalar field. We can compute the time-ordered product
\begin{equation}
\langle 0\vert T(\phi(x)\phi(x^{\prime}))\vert 0\rangle_{s}=
\sum_{\alpha} (f_{\alpha}({\bf x}) \overline{f}_{\alpha}({\bf
x}^{\prime})+f_{\alpha}({\bf x}^{\prime})
\overline{f}_{\alpha}({\bf x})) \exp(-i\omega_{\alpha}\vert
t-t^{\prime}\vert)\equiv G_{F}(x,x^{\prime}) \end{equation}
$G_{F}$ is the solution of the equation
\begin{equation}
\partial_{A}(g^{AB}\sqrt{g}\partial_{B})G_{F}=-i\delta
\end{equation}
It follows that  $\langle 0\vert T(\phi(x)\phi(x^{\prime}))\vert
0\rangle_{s}=-i \Box_{g}^{-1}(x,x^{\prime})$, where $\Box_{g}$
denotes the wave operator on a manifold. As usual for the fields
linear in creation and annihilation operators the higher order
correlation functions are expressed by the two-point function.
Then, the subsequent average over the metric concerns an average
of the propagators $G_{F}$.

We can also derive  the time evolution (3) from the canonical
formalism
\begin{displaymath}
[\phi({\bf x}),\Pi({\bf y})]=i\delta({\bf x}-{\bf y})
\end{displaymath}
and
\begin{displaymath}
H=\int d{\bf x}\sqrt{g}\Big(\frac{-g_{00}}{g} ({\bf x}) \Pi({\bf
x})\Pi({\bf x})+g^{jk}({\bf x})\partial_{j}\phi({\bf x})
\partial_{k}\phi({\bf x})\Big)
\end{displaymath}
It is understood that there is a normal ordering of the bilinear
forms in $\Pi$ and $\phi$ in $H$. The normal ordering requires
metric-dependent counterterms. Such terms would change the
gravitational dynamics but in our approximation we have to neglect
altogether the gravitational dynamics later on in order to perform
explicit calculations. The non-linear local function of the
quantum metric field in $H$ itself needs a proper definition.
 The scalar field
Hamiltonian leads to the wave equation (2) no matter what this
definition is ( after the normal ordering of the scalar fields ).

If $g^{AB}=\eta^{AB}+h^{AB}$ then $H=H_{0}+H_{I}$ where $H_{I}$ is
the interaction Hamiltonian for the computation of the $S$-matrix
(see, eq.(8) below).

The metric $h^{\mu\nu}$ may be operator-valued but we assume that
the operators $h^{\mu\nu}$ commute ($h^{jk}$ are treated as a
fixed classical background). In particular, we consider a model of
relativistic fields $h^{\mu\nu}({\bf x},t)$ such that
 $h^{\mu\nu}({\bf x})$ are time-zero fields which live in the Fock
space ${\cal F}_{gr}$ and are expanded in creation and
annihilation operators ( we have in mind a model of a canonical
free field)
\begin{equation} h^{\mu\nu}(t,{\bf
x})=\sum \Big(f_{\alpha}^{\mu\nu}({\bf
x})\exp(-i\nu_{\alpha}t)c_{\alpha}+
\exp(i\nu_{\alpha}t)\overline{f}_{\alpha}^{\mu\nu}({\bf
x})c_{\alpha}^{+}\Big)
\end{equation}
where $f^{\mu\nu}_{\alpha}({\bf x})\exp(-i\nu_{\alpha}t)$ are
solutions of the wave equation on a certain classical background .
The Hamiltonian for the metric field is
\begin{displaymath} H_{g}=
\sum_{\alpha}\nu_{\alpha}c^{+}_{\alpha}c_{\alpha}
\end{displaymath}
If the metric $h$ is defined on $R^{d}$ (a flat background metric)
then we have $\nu_{\alpha}=c\vert {\bf k}\vert$ and
 $\alpha=(\sigma,{\bf k})$ where $\sigma$ is a
 polarization index, ${\bf k}$ is the momentum and
\begin{displaymath}
f^{\mu\nu}_{\alpha}=P_{\sigma}^{\mu\nu}({\bf
k})(2\pi)^{-\frac{d}{2}}(c\vert {\bf
k}\vert)^{-\frac{1}{2}}\cos({\bf k}{\bf x})
\end{displaymath}
where $P$ depends on the choice of coordinates (the
gauge)\cite{wein}. We consider  quantum gravitons at the
temperature T  described by the Gibbs density matrix $\exp(-\beta
H_{g})$ where $\frac{1}{\beta}=k_{B}T$ and $k_{B}$ is the Boltzman
constant. Then, the time ordered product of the metric fields
reads
\begin{displaymath}
\begin{array}{l}
{\cal D}^{\mu\nu;\gamma\rho}(x-x^{\prime})\equiv\langle 0\vert
T(\exp(-\beta H_{g}) h^{\mu\nu}(x)h^{\gamma\rho}(x^{\prime}))\vert
0\rangle_{g}=\cr (2\pi)^{-d}\hbar\int d{\bf k}(c\vert {\bf
k}\vert)^{-1}{\cal P}^{\mu\nu;\gamma\rho}({\bf k})\cos({\bf
k}({\bf x}-{\bf x}^{\prime})) \exp(-ic\vert{\bf k}\vert\vert
t-t^{\prime}\vert) +\cr (2\pi)^{-d}\hbar\int d{\bf k}(c\vert {\bf
k}\vert)^{-1}{\cal P}^{\mu\nu;\gamma\rho}({\bf k})\cos({\bf
k}({\bf x}-{\bf x}^{\prime})) (\exp(c\hbar\beta\vert {\bf
k}\vert)-1)^{-1} \cos(c\vert{\bf k}\vert(
t-t^{\prime}))\end{array}
\end{displaymath}
where
\begin{displaymath}
{\cal
P}^{\mu\nu;\gamma\rho}=\sum_{\sigma}P^{\mu\nu}_{\sigma}P^{\gamma
\rho}_{\sigma}
\end{displaymath}
depends on the gauge chosen for the metric field.
 In the limit $\hbar\rightarrow 0$ (or $T\rightarrow \infty$)
we obtain the classical correlations resulting from  the classical
Gibbs distribution $\exp(-\beta\int \nabla h\nabla h)$ of the
initial values for the gravitational field. Hence, the covariance
behaves as
\begin{displaymath}
\vert \vert {\bf x}-{\bf
x}^{\prime}\vert-c(t-t^{\prime})\vert^{-d+2}
\end{displaymath}
With a general time-independent background metric we should replace $c\vert
{\bf k}\vert$ by $\nu_{\alpha}$ and  ${\cal P}\cos({\bf k}({\bf
x}-{\bf x}^{\prime}))$ by \begin{displaymath} \sum_{\alpha}
(f_{\alpha}^{\mu\nu}({\bf x}) \overline{f}_{\alpha}^{\gamma\rho}({\bf
x}^{\prime})+f_{\alpha}^{\gamma\rho}({\bf x}^{\prime})
\overline{f}_{\alpha}^{\mu\nu}({\bf x}))
\end{displaymath}
 in the formula for
the temperature Green's functions . The short distance behaviour
remains the same.

Instead of the thermodynamic equilibrium described by the Gibbs
distribution we may consider a process of dissipation
corresponding to an absorption
 of gravitational waves. Then, an absorption and emission of gravitational
 waves may lead to a time-independent
 (equilibrium) distribution of waves at large time, e.g., if the dissipation of gravitational
 energy is of the diffusive type, then
 for long times we could obtain
 \begin{displaymath}
 \frac{dg}{dt}=\Gamma\triangle g+\Theta
 \end{displaymath}
 where the $\Gamma$-term describes a dissipation. If in the
 momentum (equiv.position) space the correlations of $\Theta$   have a $\delta$
 -type distribution then the invariant measure for
 the distribution of waves is the same as the
 one for the time-zero field
 at high temperature,i.e., $\triangle^{-1}({\bf x},{\bf y})\simeq \vert {\bf x}-{\bf y}
 \vert^{-d+2}$. In any case in sec.4 we assume the
 $\vert {\bf x}-{\bf
 y}\vert^{-d+2}$
  behaviour of the metric correlations no matter
 where it comes from.

The $S$-matrix can be calculated from its generating functional in
the interaction picture
\begin{equation}
\begin{array}{l}
 \langle 0\vert T\Big(\exp(-i\int (H_{I}(s)+\int d{\bf
x}J(s,{\bf x})\phi(s,{\bf x}))ds)\Big)\vert 0\rangle \cr =\langle
0\vert \det{\cal A}^{-\frac{1}{2}}\exp(-\frac{1}{2}JG_{F}J)\vert
0\rangle_{g} \end{array}
\end{equation} In this formula $\Pi$ and
$\phi$ on the l.h.s. undergo the free Hamiltonian evolution but
(as a simplification of our model) $g$ does not change in time. In
sec.4 we calculate the expectation values of $G_{F}$ under the
assumption that $g$ is Gaussian $d=3$ and the propagator of $g$ is
$\vert {\bf x}-{\bf
 y}\vert^{-1}$. From eq.(8) it can be seen that even if
 $g$ is the free field then the expectation values
 over $g$ of the Green's functions do not reduce to Gaussian
 integrals because of the $g$-dependence of $\det{\cal A}$.
In order to calculate an average over $g$ we can apply the formula
\begin{displaymath}
\det{\cal A}^{-\frac{1}{2}}=\exp(-\frac{1}{2}Tr\ln{\cal
A})=\exp(\frac{1}{2}Tr\int dm^{2}({\cal A} +m^{2})^{-1})
\end{displaymath}
subsequently expanding  the exponential. Such an expansion
corresponds to an expansion in the number of closed scalar loops.
 \section{The
Feynman integral for the propagator} We apply a representation of
the Feynman integral by means of stochastic processes
\cite{habajp2}-\cite{habook}. For a general metric we would need a
Brownian motion on a manifold for this purpose \cite{ike}. In
order to simplify the argument we consider the wave operator on a
manifold with a particular metric in $D$ dimensions which in
$D=d+1=4$ can be related to the one called conformally static
metric in refs.\cite{orig}-\cite{kram}. We assume
$g^{jk}\sqrt{g}=\delta^{jk}$ and $g_{00}=g^{jj}$ then
\begin{displaymath}
\begin{array}{l}
-{\cal A}= \frac{1}{2}\partial_{A}(g^{AB}\sqrt{g}\partial_{B}) \cr
\equiv\frac{1}{2}\Box_{D}+\frac{1}{2} \sum \hat{h}^{\mu\nu}({\bf
x})\partial_{\mu}\partial_{\nu}\equiv
\frac{1}{2}\hat{g}^{00}\partial_{0}^{2}+\frac{1}{2}\triangle
\end{array}
\end{displaymath}
where $\hat{g}^{AB}=\sqrt{g}g^{AB}$ and
$\hat{h}^{AB}=\sqrt{g}h^{AB}$, and we split the $D$-coordinates
$x=({\bf x},X)$ into $d$ coordinates ${\bf x}$ and $D-d$
coordinates $X$. Eq.(2) in $d=3$ can also be expressed in the form
of a wave equation on a static threedimensional manifold (here
$\triangle$ denotes the threedimensional Euclidean Laplacian)
\begin{displaymath}
(\partial_{0}^{2}+\hat{g}_{00}\triangle)\phi\equiv
\Big(\partial_{0}^{2}+\hat{g}_{00}\partial_{k}(g^{kl}\sqrt{g}\partial_{l})\Big)\phi=0
\end{displaymath}

As an auxiliary tool for a calculation of the scalar propagator we
consider the Schr\"odinger-type equation for
 the Hamiltonian ${\cal A}$
\begin{equation}
i\partial_{\tau}\psi_{\tau}({\bf x},X)={\cal A}\psi_{\tau}({\bf
x},X)
\end{equation}
We can take the Fourier transform in $X$ because $g$ depends only
on spatial coordinates ${\bf x}$. Then, eq.(9) takes the form
\begin{equation}
i\partial_{\tau}\tilde{\psi}({\bf x},P)=\tilde{{\cal
A}}\tilde{\psi}({\bf x},P)
\end{equation}
where
\begin{equation}
\tilde{{\cal
A}}=-\frac{1}{2}\triangle+\frac{1}{2}P_{\mu}P_{\nu}\hat{g}^{\mu\nu}({\bf
x})
\end{equation}

 We consider the metrics $\hat{h}$ (as well as the initial states)
which are analytic functions
\begin{equation}
\hat{h}({\bf x})=\int d{\bf p}\tilde{h}({\bf p})\exp(i{\bf px})
\end{equation}
with the growth less than $\exp(\epsilon\vert{\bf z}\vert^{2})$
(with arbitrarily small $\epsilon$) for a complex ${\bf z}$. Then,
we can express the solution of eq.(9)  by means of the Feynman
integral
\begin{equation}
\psi_{\tau}({\bf x},X)=\int dP\exp(i PX) E[
M_{\tau}\tilde{\psi}({\bf x} +\lambda {\bf b}(\tau),P)]
\end{equation}
where
\begin{displaymath}
\lambda=\sqrt{i}\equiv \frac{1}{\sqrt{2}}(1+i)
\end{displaymath}
and
\begin{equation}
M_{\tau}=\exp\Big(-\frac{i}{2}P_{\mu}P_{\nu}\int_{0}^{\tau}
\hat{g}^{\mu\nu}( {\bf x}+\lambda {\bf b}(s))ds\Big)
\end{equation}

 Eq.(14) is
understood as a limit $R\rightarrow \infty$ of a regularized
expression with
\begin{equation}
\hat{g}_{R}({\bf x}+\lambda
 {\bf b}(s))=\exp(-\frac{{\bf b}(s)^{2}}{2R})\hat{g}({\bf x}+\lambda{\bf
 b}(s))
 \end{equation}
The proof of eq.(13) (for a Hamiltonian with a potential) was
given in ref.\cite{habajp2}. The operator (11) coincides with the
one for the Schr\"odinger equation with the potential
$\frac{1}{2}PgP$.

Eq.(13) can be considered as an analytic continuation of the
imaginary time version (the diffusion equation \cite{ike})
\begin{equation}
-\partial_{\tau}\psi_{\tau}({\bf x}, P)=\tilde{{\cal
A}}\psi_{\tau}({\bf x},P)
\end{equation}
However, with $\hat{g}^{00}=-1+\hat{h}^{00}$ we  would need to
restrict ourselves to $\hat{h}^{00}$ which are bounded from below
if the diffusion equation (16) is to make sense. This is the basic
reason for an analytic continuation from the Wiener to the Feynman
integral in the final formulas in this paper. In contradistinction
to the standard methods we achieve the analytic continuation in
time through an analytic continuation in space. We would not need
to require any analycity of the potential if we worked with the
imaginary time. We would obtain the formulas (13) and (14) with
$\lambda=1$. Then, an analytic continuation to complex
$\lambda=\frac{1}{\sqrt{2}}(1+i)$ is needed. Both the imaginary
time and the real time formulas follow from the Trotter product
formula . Denoting
\begin{equation}
{\cal A}=-\frac{1}{2}\triangle + V
\end{equation}
we have for the imaginary time
\begin{equation}
\exp(-\tau{\cal
A})=lim_{n\rightarrow\infty}\Big(\exp(\frac{\tau}{2n}\triangle)
\exp(-\frac{\tau}{n}V)\Big)^{n}
\end{equation}
The kernel $K$ of the r.h.s. of the formula (18) can be expressed
by the Brownian motion. In the real time case we express the
kernel (for a finite $n$ ) by means of the free propagators and
rotate the integration axis from $x$ to $\lambda x$ . Then, the
limit $n\rightarrow \infty$ can again be expressed by the Brownian
motion (as in eq.(13)). We can express the solution of eq.(13) by
means of the Feynman kernel $K$ defined by $\psi_{\tau}(x)=\int dy
K_{\tau}(x,y)\psi(y)$. We have
\cite{habajp2}\cite{habook}\cite{sim}
\begin{equation}
\begin{array}{l}
K_{\tau}(x,y)=(2\pi)^{-D+d} (2\pi
i\tau)^{-\frac{d}{2}}\exp(\frac{i}{2\tau}({\bf y}-{\bf
x})^{2})\int dP
         \exp\left(i P\left( Y-X\right)\right)\cr
         \exp(-i\frac{\tau}{2} P^{2})
 E[\exp\left(-\frac{i}{2}\int_{0}^{\tau} P_{\mu}\hat{h}^{\mu\nu}\left ({\bf
v}\left(s,{\bf x},{\bf y}\right)\right)P_{\nu}ds\right)]
\end{array}
\end{equation}
where
\begin{equation}
{\bf v}(s,{\bf x},{\bf y})={\bf x}+\frac{s}{\tau}({\bf y}-{\bf x})
+\lambda\sqrt{\tau}{\bf a}(\frac{s}{\tau})
\end{equation}
and ${\bf a}$ is the Brownian bridge starting from $0$ and ending
at $s=\tau$ in $0$. The Brownian bridge is  defined as the
Gaussian process with mean equal to zero and the  covariance
\begin{equation}
E[a_{j}(s)a_{k}(s^{\prime})]=\delta_{jk}s(1-s^{\prime})
\end{equation}
for $s\leq s^{\prime}$. The representation \cite{sim} ${\bf
a}(s)=(1-s){\bf b}(\frac{s}{1-s})$ (where ${\bf b}$ is the
Brownian motion) is useful for computations.

\section{An average over the quantum metric}
We are interested in  metrics which are relativistic quantum
fields. In the real-time quantum field theory the time-ordered
products have the K\"allen-Lehmann representation
\cite{sch}\begin{equation} \begin{array}{l} {\cal
D}^{\mu\nu;\alpha\beta}(x,y)=\langle 0\vert
T(h^{\mu\nu}(x)h^{\alpha\beta}(y))\vert 0\rangle=\int
d\rho(m^{2}){\cal D}^{\mu\nu;\alpha\beta}_{F}(x-y;m^{2}) \cr
=lim_{\epsilon\rightarrow 0}{\cal P}^{\mu\nu;\alpha\beta}\int
d\rho(m^{2})\int_{0}^{\infty} id\tau (2\pi i
\tau)^{-\frac{d}{2}}\exp(-\frac{i}{2}m^{2}\tau-\epsilon\tau)\exp(\frac{i}{2\tau}(x-y)^{2})
\end{array}
 \end{equation}
 Hence,
\begin{equation}
\begin{array}{l}
{\cal D}_{F}(x-y)=\int d\rho(m^{2}){\cal D}_{F}(x-y;m^{2}) \cr
=lim_{\epsilon\rightarrow 0}\int_{0}^{\infty} id\tau
\sigma(\tau)\exp(-\epsilon\tau)\exp(\frac{i}{2\tau}(x-y)^{2})
\end{array}
 \end{equation}
where \begin{equation} \sigma(\tau)=\int
d\rho(m^{2})(2i\pi\tau)^{-\frac{d}{2}} \exp(-\frac{i}{2}m^{2}\tau)
\end{equation}
The Euclidean version of eq.(23) reads
\begin{equation}
\begin{array}{l}
{\cal D}({\bf x}-{\bf y})=\int_{0}^{\infty} d\tau
\sigma(-i\tau)\exp(-\frac{1}{2\tau}({\bf x}-{\bf y})^{2})
\end{array}
 \end{equation}
It can be seen that if
\begin{displaymath}
{\bf x}-{\bf y}\rightarrow {\bf x}-{\bf y}+{\bf z}
\end{displaymath}
then we can continue to complex ${\bf z}$ if
\begin{displaymath}
\Re(({\bf x}-{\bf y}+{\bf z})^{2})\geq0
\end{displaymath}
We can formulate the analycity also in terms of the radial
analycity meaning that ${\bf x}-{\bf y}\rightarrow \lambda ({\bf
x}-{\bf y})$ if $\Im(\lambda^{2}({\bf x}-{\bf y})^{2})\geq 0$.
From the point of view of the proper-time representation in the
imaginary time formulation it is useful to introduce the
regularized imaginary time metric fields in the form
\begin{equation}
h_{\delta}({\bf
x})=\int_{\delta}^{\infty}d\tau\exp(-\frac{\tau}{4}{\bf k}^{2})
\chi({\bf k},\tau)\sqrt{\sigma(-i\tau)}\exp(i{\bf k x})
\end{equation}
where \begin{equation}
 \langle\tilde{\chi}^{\mu\nu}({\bf
k},\tau)\tilde{\chi}^{\rho\sigma}({\bf
k}^{\prime},\tau^{\prime})\rangle= {\cal P}^{\mu\nu;\rho\sigma}
\delta({\bf k}+{\bf k}^{\prime})\delta(\tau-\tau^{\prime})
\end{equation}
Then, $h_{\delta}({\bf x})$ is an analytic function of ${\bf x}$
and its correlation function is ${\cal D}_{\delta}$ where in
eq.(25) the $\tau$-integral starts from $\tau \geq \delta$. For
further computations we assume that the metric is Gaussian. We
have seen in eq.(8) that an average over the scalar field  leads
to a determinant of ${\cal A}$ which depends on the metric $g$. It
is difficult to calculate non-perturbatively a contribution of the
non-Gaussian terms in the complete theory of the interaction of
gravity with a scalar field. We are able to calculate the averages
over
 the metric only through an expansion in the number of closed scalar
 loops as discussed at the end of sec.2.
 In such an expansion the ultraviolet behaviour does not depend
 on the number of loops. Hence, we may restrict ourselves here to the
 zeroth order corresponding to $\det {\cal A}=1$.

Using the Trotter product formula (18) (with $V=\frac{1}{2}PgP$
and $\tau \rightarrow -i\tau$) we obtain (for a regular Gaussian
metric (26), see eq.(8) for a calculation of the Gaussian average)
\begin{equation}
\begin{array}{l}
\langle 0\vert
\Big(\exp(\frac{i\tau}{2n}\triangle)\exp(-\frac{i\tau}{n}V)\Big)^{n}
\vert 0\rangle_{gr}(X, {\bf x};Y, {\bf y}) \cr =\int
dP\exp(iP(X-Y))(2\pi
i\frac{\tau}{n})^{-\frac{d}{2}}\exp(in\frac{({\bf x}-{\bf
x}_{1})^{2}}{2\tau})

\cr .....\exp\Big(-\frac{\tau^{2}}{8n^{2}}\sum PP {\cal
D}_{\delta}({\bf x}_{j}-{\bf x}_{j-1})PP\Big) d{\bf
x}_{1}....d{\bf x}_{n}
\end{array}
\end{equation}
In this equation we rotate the integration line ${\bf
x}\rightarrow \lambda {\bf x}$. Then, the limit $n\rightarrow
\infty$ can be expressed by the Brownian motion. As applied to
propagators instead of the kernel ${\cal A}^{-1}(x,y)$ consider
${\cal A}^{-1}(\lambda {\bf x},X;\lambda {\bf y},Y)$. Using the
Brownian bridge representation we obtain for an average (28) over
the metric (22)
\begin{equation}
\begin{array}{l}
\langle {\cal A}^{-1}(\lambda {\bf x},X;\lambda {\bf y},Y)\rangle
=\int_{0}^{\infty} d\tau\exp( iP( X- Y))
(2\pi\tau)^{-\frac{d}{2}}\exp(-\frac{1}{2\tau}({\bf x}-{\bf
y})^{2}) \cr E[\exp\Big(
-\frac{1}{8}\int_{0}^{\tau}\int_{0}^{\tau}dsds^{\prime}PP {\cal
D}_{\delta}(\lambda(\tilde{{\bf v}}(s)-\tilde{{\bf
v}}(s^{\prime}))PP\Big) ]
\end{array}
\end{equation}
where
\begin{displaymath}
\tilde{{\bf v}}(s)={\bf x}+\frac{s}{\tau}({\bf y}-{\bf
x})+\sqrt{\tau}{\bf a}(\frac{s}{\tau})
\end{displaymath}

 We can calculate higher order correlations
\begin{equation}
\begin{array}{l}
\langle {\cal A}^{-1}(\lambda {\bf x},X;\lambda {\bf y},Y){\cal
A}^{-1}(\lambda {\bf x}^{\prime},X^{\prime};\lambda {\bf y}
^{\prime},Y^{\prime})\rangle =\int_{0}^{\infty}
d\tau_{1}\int_{0}^{\infty} d\tau_{2}d Pd P^{\prime} \cr \exp( i P(
X- Y)) (2\pi\tau_{1})^{-\frac{d}{2}}\exp(-\frac{1}{2\tau_{1}}({\bf
x}-{\bf y})^{2})\exp( i P^{\prime}( X^{\prime}- Y^{\prime}))
(2\pi\tau_{2})^{-\frac{d}{2}} \cr \exp(-\frac{1}{2\tau_{2}}({\bf
x}^{\prime}-{\bf y}^{\prime})^{2}) \cr E[\exp\Big(
-\frac{1}{8}\int_{0}^{\tau_{1}}\int_{0}^{\tau_{1}

}dsds^{\prime}PP {\cal D}_{\delta}(\lambda(\tilde{{\bf
v}}(s)-\tilde{{\bf v}}(s^{\prime}))PP \cr
-\frac{1}{8}\int_{0}^{\tau_{2}}\int_{0}^{\tau_{2}}dsds^{\prime}P^{\prime}P^{\prime}
{\cal D}_{\delta}(\lambda(\tilde{{\bf v}}^{\prime}(s)-\tilde{{\bf
v}}^{\prime}(s^{\prime}))P^{\prime}P^{\prime} \cr
+\frac{1}{4}\int_{0}^{\tau_{2}}\int_{0}^{\tau_{1}}dsds^{\prime}PP
{\cal D}_{\delta}(\lambda(\tilde{{\bf v}}^{\prime}(s)-\tilde{{\bf
v}}(s^{\prime}))P^{\prime}P^{\prime}\Big) ] \cr +(x\rightarrow
x^{\prime})\end{array}
\end{equation}
Here, $(x\rightarrow x^{\prime})$ means the same expression with
$x$ exchanged with $x^{\prime}$, $\tilde{{\bf v}}^{\prime}$ is an
independent bridge with $({\bf x},{\bf
y})\rightarrow ({\bf x}^{\prime},{\bf y}^{\prime})$. In
ref.\cite{haba3} we have shown that the formulas (29)-(30) come
from a resummation of the perturbation series in $\hat{h}$ of
$\Box_{g}^{-1}$. It follows from eq.(8) that this is also a
resummation of the Dyson series in the interaction Hamiltonian. We
could derive the short distance behaviour of the $4$-point
function from eq.(30). It can be shown \cite{habajp} that its
singularity is a product of the singularities of the two-point
functions.

Whether the correlations (29)-(30) are finite or not depends on
whether the argument in the exponent is a well-defined function
and whether the exponential factors are integrable. The functions
in the exponent are analytic continuations from real arguments. It
will be useful to use a concrete representation of these
functions. So, for our special case  of the $\vert {\bf x} -{\bf
y}\vert^{-1}$ metric correlations we consider the representation
\begin{equation}
(\frac{1}{\vert{\bf
k}\vert})^{2}=\int_{0}^{\infty}dr\exp(-\frac{1}{2}r\vert {\bf
k}\vert^{2})\Big(\int_{0}^{\infty} dr\exp(-\frac{1}{2}r)
\Big)^{-1}
\end{equation}
Then, we can perform the integral over ${\bf k}$ (with the
regularization (26))\begin{equation}\begin{array}{l} {\cal
D}_{\delta}({\bf x}-{\bf y})= \int_{\delta}^{\infty}dr\int d{\bf
k}\exp(-\frac{1}{2}r \vert {\bf k}\vert^{2})\exp(i{\bf k}({\bf
x}-{\bf y}))= \cr \int_{\delta}^{\infty}dr(2\pi
r)^{-\frac{d}{2}}\exp\Big(-\frac{1}{2} r^{-1}({\bf x}-{\bf
y})^{2}\Big)
\end{array}
\end{equation}
From the representation (32) it can be seen that this is an
analytic function of ${\bf z}^{2}=({\bf x}-{\bf y})^{2}$ as long
as $\Re {\bf z}^{2}\geq 0$ (the equality sign is allowed only if
$\Im {\bf z}^{2}\neq 0$). The argument of the exponential factors
in eqs.(29)-(30) is a well-defined random variable  because it is
square integrable. In fact, the expectation value
\begin{equation}
\begin{array}{l}
E[\vert\int ds ds^{\prime}{\cal D}_{\delta}(\lambda(\tilde{{\bf
v}}(s)-\tilde{{\bf v}}(s^{\prime}))\vert^{2}] \cr =\int ds
ds^{\prime}dtdt^{\prime}E[\overline{{\cal D}_{\delta}(\lambda(\tilde{{\bf
v}}(s)-\tilde{{\bf v}}(s^{\prime}))}{\cal D}_{\delta}(\lambda(\tilde{{\bf
v}}(t)-\tilde{{\bf v}}(t^{\prime}))]
\end{array}
\end{equation}
can be explicitly calculated (using eq.(21)) and shown to be
finite for  $\delta >0$ as well as in the limit $\delta
\rightarrow 0$. When $\delta =0$ the complex argument in the
exponent in eqs.(29)-(30)) for the propagator (32) scales as
\begin{equation}
-{\cal D}(\lambda(\tilde{{\bf v}}(s)-\tilde{{\bf
v}}(s^{\prime})))=-\frac{1}{\lambda}{\cal D}(\tilde{{\bf v}}(s)-\tilde{{\bf
v}}(s^{\prime})) \end{equation} where ${\cal D}$ is positive.
Hence, the real part of (34) is negative supplying a damping
factor for the integrals of the exponential factors (29)-(30).

After the proof that the expressions (29)-(30) are finite we
obtain  just by scaling ( as in \cite{habajp}, although
$g^{\mu\nu}=\eta^{\mu\nu} + h^{\mu\nu}$ is not scale invariant)
  \begin{equation}
  \langle {\cal A}^{-1}(X,Y)\rangle\simeq  \vert
   X- Y\vert^{-2-\frac{1}{3}}
  \end{equation}
  at short distances (and in the approximation $\det {\cal A}=1$).

   For the behaviour in the ${\bf x}$ direction we put $X=Y=0$
 then
   \begin{equation}
  \langle {\cal A}^{-1}({\bf x},{\bf y})\rangle=  R  \vert
  {\bf x}-{\bf y}\vert^{-2+\frac{1}{4}}
  \end{equation}
  where $R$ is a  constant. It is more regular than the two-point function
  for the
   four-dimensional free field (see ref.\cite{PLB}
  for a more general argument). In order to treat the general case
  of
  $2n$-point functions and $\det {\cal A}\neq 1$ we expand $\det{\cal A}$
  in eq.(8) in terms of the Green's functions as discussed at the
  end of sec.2. Then, at each order we can calculate the
  expectation values of products of Green's functions (as in
  eq.(30), see also \cite{habajp}).
  In each order of the expansion of $\det{\cal A}$  the results (35)-(36) can be generalized to
  arbitrary $2n$-point correlation functions proving
  their anomalous short distance behaviour.
   \section{Discussion}
We have discussed a Hamiltonian formulation of an interaction of a
quantum metric with a quantum scalar field. We have compared the
standard Dyson expansion for this model with the results coming
from the functional integral. We have calculated scalar field
correlation functions in the Gaussian approximation for the metric
field expanding the correlation functions in the number of closed
scalar loops. The ultraviolet behaviour does not depend on the
number of loops. We have shown that at each order of the expansion
the short distance behaviour is substantially modified as a result
of the interaction with gravitons. In particular, the behaviour at
equal times of the correlation functions is more regular than the
one for the free scalar field. In sec.4 we concentrated on the
static metric corresponding to $D=d+1=4$ of the  framework of
refs.\cite{hab1}-\cite{habajmp} because it may be relevant to
physical models of gravitons in equilibrium with matter. The
Hamiltonian framework of this paper may be considered for any
$d<D$. From the S-matrix formula  we can obtain (inserting static
sources in eq.(8); see also a more detailed discussion in
\cite{sch} ) the expression for the Yukawa potential
\begin{displaymath}
V({\bf x},\tilde{{\bf X}})=\int_{-\infty}^{\infty}dX_{D}\langle
{\cal A}^{-1}({\bf x},\tilde{{\bf X}},X_{D})\rangle
\end{displaymath}
here we denoted $X=(\tilde{{\bf X}},X_{D})$, where $X_{D}$ is the
time. If ${\cal D}({\bf x})\simeq \vert {\bf x}\vert^{-4\gamma}$
at short distances then using just scaling properties of the
formula (29) (as in refs.\cite{hab1}-\cite{habajmp}) we can
conclude that
\begin{displaymath}
V({\bf x},\tilde{\bf X}=0)\simeq \vert {\bf
x}\vert^{2-d-(1-\gamma)(D-d-1)}
\end{displaymath}
Hence, in general the Yukawa potential is less singular in the
${\bf x}$ direction than the canonical one ( except the static
case of sec.4 corresponding to $D=d+1$ when the Yukawa potential
does not change in the ${\bf x}$ direction at short distances).

In the $\tilde{{\bf X}} $ direction (if $D>d+1$)
\begin{displaymath}
V({\bf x}=0,{\tilde {\bf X}})=\vert \tilde{{\bf
X}}\vert^{-D+\frac{3}{1-\gamma}-(d+1)\frac{\gamma}{1-\gamma}}
\end{displaymath}
The distance scale at which the anomalous behaviour would appear
is determined by the scaling behaviour of the metric field. In
Einstein gravity there is a dimensional parameter of the Planck
scale. In this case the eventual change of the short distance
behaviour is expected below this length scale.

\end{document}